\documentclass[12pt]{article}

\usepackage[a4paper,left=2cm,right=2cm,top=2.5cm,bottom=3.5cm]{geometry}

\usepackage{amssymb} 
\usepackage{amsbsy}
\usepackage{amsmath}
\usepackage{graphicx}
\usepackage{booktabs}       % professional-quality tables
\usepackage{cite}
\usepackage{xcolor}

\usepackage[%
pdfsubject={Complex Langevin},
pdfkeywords={CL},
colorlinks=true,    %color the links, not boxes
citecolor=blue,     %Change colors of citations
linkcolor=black,    %All internal links. contents page, etc %Eqn/fig set in main.tex hypersetup
urlcolor=blue,      %Color of URLs in bib
]{hyperref}

\newcommand{\el}{\vspace*{0.3cm}}
\newcommand{\be}{\begin{equation}}
\newcommand{\ee}{\end{equation}}

\newcommand{\cN}{{\cal N}}

\newcommand{\E}{\mathbb{E}}
\newcommand{\half}{\frac{1}{2}}
\newcommand{\ket}{\right\rangle}
\newcommand{\bra}{\left\langle}
\newcommand{\qqquad}{\qquad\quad}

\newcommand{\rv}{{\mathbf r}}
\newcommand{\sv}{{\mathbf s}}
\newcommand{\vv}{{\mathbf v}}
\newcommand{\xv}{{\mathbf x}}
\newcommand{\Kv}{{\mathbf K}}
\newcommand{\Av}{{\mathbf A}}
\renewcommand{\Re}{\mbox{Re}\,}
\renewcommand{\Im}{\mbox{Im}\,}

\newcommand{\eps}{\epsilon}

\newcommand{\bean}{\begin{eqnarray*}}
\newcommand{\eean}{\end{eqnarray*}}

\newcommand{\vecnul}{{\mathbf 0}}
\newcommand{\etav}{{\boldsymbol\eta}}

\newcommand{\Id}{1\!\!1}

\newcommand{\bignorm}{\big|\big|}

\renewcommand{\theequation}{\arabic{section}.\arabic{equation}}
\counterwithin*{equation}{section}

\begin{document}

\title{\bf\large
Combining complex Langevin dynamics with score-based and energy-based diffusion models
}

\author{
 Gert Aarts,$^{a}$ 
 Diaa E.\ Habibi,$^{a}$ 
 Lingxiao Wang$^{b}$ 
 and Kai Zhou$^{c,d}$
 \footnote{
 email:  \{g.aarts, n.e.habibi\}@swansea.ac.uk, lingxiao.wang@riken.jp,  zhoukai@cuhk.edu.cn}
 \\
\mbox{} \\
{\small 
${}^a$Centre for Quantum Fields and Gravity} \\
{\small Department of Physics, Swansea University, Swansea SA2 8PP,  United Kingdom} \\
{\small
${}^b$RIKEN Center for Interdisciplinary Theoretical and Mathematical Sciences (iTHEMS)}\\
{\small
Wako, Saitama 351-0198, Japan} \\
{\small
${}^c$School of Science and Engineering, The Chinese University of Hong Kong} \\
{\small
Shenzhen (CUHK-Shenzhen), Guangdong, 518172, China} \\
{\small
${}^d$Frankfurt Institute for Advanced Studies, Ruth Moufang Strasse 1} \\
{\small 
D-60438, Frankfurt am Main, Germany} \\
}

\date{\normalsize October 1, 2025}

\maketitle

\vspace*{-0.5cm}

\begin{abstract}
Theories with a sign problem due to a complex action or Boltzmann weight can sometimes be numerically solved using a stochastic process in the complexified configuration space. However, the probability distribution effectively sampled by this complex Langevin process is not known a priori and notoriously hard to understand. In generative AI, diffusion models can learn distributions, or their log derivatives, from data. We explore the ability of diffusion models to learn the distributions sampled by a complex Langevin process, comparing score-based and energy-based diffusion models, and speculate about possible applications. 
\end{abstract}

%%%%%%%%%%%%%%%%%%%%%%%%%%%%%%%%%%%%%%%%%%%%%%%%%%%

\newpage

\tableofcontents

\hypersetup{linkcolor=blue}   %All internal links

\section{Introduction}

Theories with a complex Boltzmann weight suffer from sign and overlap problems, which makes them hard to simulate using conventional numerical methods based on importance sampling \cite{Troyer:2004ge}. 
A well-known example is QCD at non-zero baryon density, for which  the quark determinant is complex-valued when the quark chemical potential is real,
\be
\left[\det M(\mu)\right]^* = \det M(-\mu^*) \in \mathbb C,
\ee
see e.g.~the reviews \cite{deForcrand:2009zkb,Aarts:2015tyj}. A potential solution is to take the complexity seriously and extend the theory into the complex plane, or complexified manifold more generally. This idea underpins complex Langevin dynamics \cite{Parisi:1983mgm,klauder}, Lefschetz thimbles \cite{Cristoforetti:2012su}, holomomorphic flow \cite{Alexandru:2015sua}, and variations thereof. 

Complex Langevin (CL) dynamics, in which the degrees of freedom are analytically extended, does not rely on importance sampling but explores a complexified manifold via a stochastic process \cite{Parisi:1983mgm,klauder}. It is an extension of stochastic quantisation \cite{Parisi:1980ys,Damgaard:1987rr}.
CL has been shown to work in lattice field theories in three \cite{Aarts:2011zn} and four \cite{Aarts:2008wh} Euclidean dimensions with a severe sign problem, including in QCD in four dimensions \cite{Seiler:2012wz,Sexty:2013ica,Aarts:2016qrv,Sexty:2019vqx,Scherzer:2020kiu,Ito:2020mys,Tsutsui:2025jez}, but it may also fail, even in simple models 
\cite{Ambjorn:1985iw,Aarts:2010aq,Aarts:2012ft}.
This situation was clarified a few years ago \cite{Aarts:2009uq,Aarts:2011ax,Nishimura:2015pba}
by the derivation of the formal relation between the complex distribution on the real manifold and the real and semi-positive distribution on the complexified manifold, which is effectively sampled during the CL process. This resulted in practical criteria for correctness which need to be verified {\em a posteriori} \cite{Aarts:2009uq,Aarts:2011ax,Nishimura:2015pba}. Nevertheless, issues remain and the reliability of the method depends on a precise understanding of the behaviour of this distribution, in particular at infinity and near poles in the CL drift. Recent work can be found in e.g.\ Refs.~\cite{Scherzer:2018hid,Scherzer:2019lrh,Alvestad:2022abf,Seiler:2023kes,
Hansen:2024lkn,Boguslavski:2024yto,Mandl:2025mav}.

A crucial role is played by the distribution on the complexified manifold.
Unfortunately, this distribution turns out to be elusive, as the Fokker-Planck equation linked to the CL process cannot be solved in general. In fact, even convergence is hard to understand, except in some simple cases, such as Gaussian models \cite{Aarts:2009hn} and models in which the decay at infinity can be precisely understood (see below) \cite{Aarts:2013uza}. A better characterisation of the distribution would, therefore, be welcome. 
Independently of which approach is used, relating averages over complex-valued Boltzmann weights to statistical averages over real, semi-positive probability distributions is an open problem worth studying more generally \cite{Weingarten:2002xs,Salcedo:2018fvt}.

Diffusion models \cite{sohl-dickstein:2015deep,ho:2020denoising,SongErmon2020:generativemodels,Song:2021scorebased,yang:2022diffusion,song2021maximumlikelihoodtrainingscorebased,karras2022elucidatingdesignspacediffusionbased} 
are a class of generative methods, which learn distributions from data. 
They rely on a stochastic process, similar to stochastic quantisation, but instead of using a known drift term derived from the underlying theory, they learn the drift from data previously collected. We have recently explored the relation between diffusion models and stochastic quantisation in scalar \cite{Wang:2023exq,Wang:2023sry} and U(1) gauge theories \cite{Zhu:2025pmw}, and studied the evolution of higher-order cumulants in detail \cite{Aarts:2024rsl}. Further connections between diffusion models, field theory and the renormalisation group are pointed out in Refs.~\cite{cotler2023renormalizingdiffusionmodels,Hirono:2024zyg,Fukushima:2024oij}.
Diffusion models are very flexible and can be formulated e.g.\ in real space or momentum space \cite{Gerdes:2024iyg}. 

Since diffusion models do not require {\em a priori} knowledge of the distribution itself but learn directly from data, they can be employed to study the distribution sampled during the CL process \cite{Habibi:2024fbn}. When successfully trained, diffusion models can then be used to both generate additional configurations and to deepen our understanding of the real distribution on the complexified manifold, offering a fresh perspective. This is what we explore here.

The paper is organised as follows. We start with a brief summary of CL dynamics in Sec.~\ref{sec:cl}. In Sec.~\ref{sec:dm}, we first introduce diffusion models and then discuss score-based and energy-based diffusion models to construct distributions and their log derivatives. Since the purpose of this paper is to address conceptual questions, we apply the framework in Sec.~\ref{sec:quartic} to a simple and well-understood model \cite{Aarts:2013uza}. A summary and outlook are given in Sec.~\ref{sec:out}. Appendix~\ref{sec:comp} contains computational details. To ensure that our findings are not specific to the complexification required for CL dynamics, we study in Appendix~\ref{sec:gaussian} an exactly solvable model with a real distribution to start with. Appendix~\ref{sec:thimbles} contains a comparison with Lefschetz thimbles. Appendix~\ref{sec:system} finally contains our approach to estimating systematic uncertainties.

\section{Complex Langevin dynamics}
\label{sec:cl}

We start by summarising the idea behind stochastic quantisation and complex Langevin dynamics, considering for simplicity one degree of freedom $x$ with a complex-valued Boltzmann weight $\rho(x)$. The setup is easily extended to many degrees of freedom and (lattice) field theory \cite{Parisi:1983mgm,Berges:2007nr,Aarts:2008rr,Sexty:2013ica}.

As always, observables are defined as 
\be
\bra O(x)\ket = \int dx\, \rho(x)O(x), \qqquad \rho(x) = \frac{1}{Z} \exp[-S(x)], \qqquad Z = \int dx\,\rho(x).
\ee
The Langevin process and drift read
\be
\label{eq:L}
\dot x(t) = K[x(t)] +\sqrt{2}\eta(t), \qqquad K(x) = \frac{d}{dx} \log\rho(x) = -\frac{dS(x)}{dx},
\ee
where the dot indicates the Langevin time derivative and the noise $\eta\sim\mathcal{N}(0,1)$ is Gaussian.
The factor of $\sqrt{2}$ in the noise term is conventional and can be exchanged for a factor of 1/2 in front of the drift, by rescaling Langevin time.   
The corresponding Fokker-Planck equation (FPE) is
\be
\partial_t \rho(x;t) = \partial_x\left[ \partial_x -K(x) \right] \rho(x;t).
\ee
For a real Boltzmann weight and drift, this process converges to the stationary solution $\rho(x)\sim \exp[-S(x)]$, typically exponentially fast \cite{Damgaard:1987rr}.

When the weight is complex, one may extend $x\to z=x+iy$ into the complex plane and write
\be
\dot z(t) = K[z(t)] + \sqrt{2} \eta(t), \qqquad K(z) = \frac{d}{dz} \log\rho(z) = -\frac{dS(z)}{dz}.
\ee
However, in this case the FPE cannot be used to show convergence as the corresponding Fokker-Planck Hamiltonian is no longer semi-positive definite \cite{Damgaard:1987rr}. 

Instead, one may take the real and imaginary parts of the equation above and consider the CL process,
\be
\label{eq:CL}
\begin{aligned}
\dot x(t) = K_x[x(t), y(t)] + \sqrt{2N_x} \eta_x(t), & \qqquad K_x(x,y) = \Re \frac{d}{dz} \log\rho(z) \Big|_{z\to x+iy},\\
\dot y(t) = K_y[x(t), y(t)] + \sqrt{2N_y} \eta_y(t),  & \qqquad K_y(x,y) = \Im \frac{d}{dz} \log\rho(z)\Big|_{z\to x+iy},
\end{aligned}
\ee
with the constraint $N_x-N_y=1$.
The FPE for this process reads
\be
\label{eq:FPE}
\partial_t p(x,y;t) = \left[\partial_x\left(N_x\partial_x - K_x\right) + \partial_y\left(N_y\partial_y - K_y\right) \right] p(x,y;t),
\ee
such that
\be
\bra O[x(t)+iy(t)]\ket_\eta = \int dxdy\, p(x,y;t)O(x+iy).
\ee
It is preferable to consider real noise, $N_x=1, N_y=0$ \cite{Aarts:2009uq}. Unlike the original weight $\rho(z)$, $p(x,y;t)$ is real and semi-positive definite, as it represents the distribution effectively sampled by the real Langevin process in the two-dimensional plane.

This setup should be contrasted with the case of a proper two-dimensional process starting from a known (real and semi-positive) distribution $p(\xv)$, with $\xv=(x,y)$. In that case, one can apply real Langevin dynamics, given by
\be
\label{eq:realL}
\dot \xv(t) = \Kv[\xv(t)] + \sqrt{2} \etav(t), \qqquad \Kv(\xv) = \nabla \log p(\xv), \qquad \bra\eta_i(t)\eta_j(t')\ket=\delta_{ij}\delta(t-t'),
\ee
with the corresponding FPE,
\be
\partial_t p(\xv;t) = \nabla\cdot \left[ \nabla  - \Kv(\xv)\right] p(\xv;t).
\ee
It is easy to see that $p(\xv)$ is a stationary solution of this FPE. 

Going back to the CL equation, this process yields the correct answer if a stationary solution to the FPE (\ref{eq:FPE}) exists, such that \cite{Aarts:2009uq,Aarts:2011ax}
\be
\int dxdy\, p(x,y)O(x+iy) = \int dx\, \rho(x) O(x),
\ee
or, shifting the integration variables at a formal level, 
\be
\label{eq:rhoP}
\rho(x) = \int dy\, p(x-iy,y).
\ee
Considerable effort has been invested in deriving criteria for correctness related to the behaviour of $p(x,y)$ at infinity and near poles of the drift (if there are any), which can be used a posteriori to justify the results \cite{Aarts:2009uq,Aarts:2011ax,Nishimura:2015pba,Scherzer:2018hid,Scherzer:2019lrh,Alvestad:2022abf,Seiler:2023kes,Hansen:2024lkn,Boguslavski:2024yto, Mandl:2025mav}. A better understanding of $p(x,y)$ in the stationary limit is very welcome. We turn to diffusion models to help with this question.

\section{Score and energy-based diffusion models}
\label{sec:dm}

Diffusion models are a class of probabilistic generative models, which learn from a data set representing the target distribution $p_0(\xv)$ \cite{sohl-dickstein:2015deep,ho:2020denoising,SongErmon2020:generativemodels,Song:2021scorebased,yang:2022diffusion}.
During the forward process, data are gradually corrupted, by applying noise which increases incrementally in strength, and a neural network is trained to learn the change in the probability distribution, using the decomposition
\be
\label{eq:ppp}
p(\xv,t) = \int d\xv_0\, p(\xv,t|\xv_0)p_0(\xv_0).
\ee
The process can then be reversed and, after training, new instances representative of the original data set are generated during the backward process.

We use the description in terms of stochastic differential equations (SDEs) \cite{Song:2021scorebased,song2021maximumlikelihoodtrainingscorebased,
yang:2022diffusion,karras2022elucidatingdesignspacediffusionbased} and refer to Refs.~\cite{Wang:2023exq,Wang:2023sry,Zhu:2025pmw} for applications in lattice field theory. 
Ref.~\cite{Aarts:2024rsl} contains a detailed analysis of the evolution of higher-order cumulants. 
The forward process is described by the SDE
\be
\dot{\xv}(t) = \half\Kv[\xv(t), t] + g(t)\etav(t),
\label{eq:forward_process}
\ee
where $\Kv[x(t), t]$ is a possible drift term and $\etav\sim\mathcal{N}(0,1)$ is again Gaussian noise. 
Compared to the previous section, we have rescaled time with a factor of 2, see Eqs.~(\ref{eq:L}, \ref{eq:realL}), and introduced a diffusion coefficient $g(t)$, setting the time-dependent noise strength.
The initial conditions for this process are determined by the target distribution $\xv(0) = \xv_0 \sim p_0(\xv_0)$ and the process runs between $0\leq t \leq T$. 

The corresponding SDE for the denoising or backward process reads \cite{anderson:1982reversetime}
\be
\dot \xv(t) = \half\Kv[\xv(t), t] - g^2(t)\nabla \log p(\xv, t) + g(t)\etav(t),
\label{eq:backward_process}
\ee
where the process now starts at $t=T$ and time runs backwards to $t=0$.
Initial conditions are sampled from a simple prior distribution, such as the normal distribution with a variance comparable to the variance obtained at the end of the forward process. 
The additional term in the drift, the so-called score, $\nabla \log p(\xv,t)$, is not known {\em a priori} and is modelled by a neural network $\sv_\theta(\xv,t)$, where $\theta$ denotes all the trainable network parameters. It is learned during the forward process, by minimising a loss function imposing
\be
\label{eq:s}
\sv_\theta(\xv,t) \approx \nabla \log p(\xv, t),
\ee
in conjunction with the decomposition (\ref{eq:ppp}) and Jensen's inequality.
After the diffusion model has been trained, new samples from the target distribution can be generated by numerically solving the backward process (\ref{eq:backward_process}) using Eq.~(\ref{eq:s}),
\be
\dot \xv(t) = \half\Kv[\xv(t), t] - g^2(t) \sv_{\theta^*}(\xv,t) + g(t)\etav(t),
\label{eq:backward_process_s}
\ee
at a (near-)optimal set of network parameters $\theta^*$.
When the target distribution is known, it is possible to include an accept/reject step \cite{Zhu:2025pmw}.
The similarities between stochastic quantisation and the dynamics in diffusion models have been noted \cite{Wang:2023exq}.

\subsection{Score-based models}

Starting from the Fisher divergence \cite{Song:2021scorebased}, 
 \be
 \mathcal{F}(\theta, \lambda) = \frac{1}{2}\int_0^T dt\, \E_{p(\xv,t)}\left[\lambda(t) \bignorm \sv_\theta(\xv,t) - \nabla\log p(\xv,t)\bignorm^2\right],
  \label{eq:fisher}
\ee
 score-based models are trained using the loss function 
 \be
 \mathcal{L}(\theta, \lambda) = \frac{1}{2}\int_0^T dt\, \E_{p(\xv,t)}\left[\lambda(t) \bignorm \sv_\theta(\xv,t) - \nabla\log p(\xv,t|\xv_0)\bignorm^2\right],
   \label{eq:loss}
 \ee
 where the weight $\lambda(t)$ is chosen to be the variance of the noise at time $t$ and Jensen's inequality was used in combination with Eq.~(\ref{eq:ppp}). The expectation includes the average over the initial distribution $p_0(\xv)$, obtained by summing over the data set. 

Given Eqs.~(\ref{eq:s}, \ref{eq:fisher}), it is tempting to identify the integrated score at the end of the backward process with the log of the target distribution function. This identification would provide a handle on the (unnormalised) data distribution, which may be especially useful for complex Langevin dynamics, in which this distribution is elusive, as mentioned above. However, here we note (as has also been observed before, see e.g.\ Refs.~\cite{saremi2019approximatingnablafneural,shouldEBMs, horvat2024gaugefreedomconservativityintrinsic}) that there is no guarantee that the score is conservative. Indeed, as we will see below, in general one finds that the score contains both a conservative and non-conservative component,
\be
\label{eq:decomp}
\sv_\theta(\xv,t) = \nabla\Phi_\theta(\xv,t) + \rv_\theta(\xv,t), \qqquad \nabla\cdot \rv_\theta(\xv,t)=0,
\ee
and is hence not integrable. Note that this is not in conflict with the learning objective. Inserting the decomposition (\ref{eq:decomp}) in Eq.~(\ref{eq:fisher}), one finds for the norm under the integral
\be
\bignorm \sv_\theta(\xv,t) - \nabla\log p(\xv,t)\bignorm^2 = 
\bignorm \nabla\Phi_\theta(\xv,t) - \nabla\log p(\xv,t)\bignorm^2
+
\bignorm \rv_\theta(\xv,t) \bignorm^2,
\ee
where we used partial integration in the loss function, ignoring boundary terms. 
Hence the (non)conservative parts are independently minimised.
However, due to the non-conservative component, the integrated score,
\be
\Phi_\theta(\xv, t) = \Phi_\theta(\vecnul,t) + \int_{\gamma:\vecnul\to\xv} d\xv'\cdot \sv_\theta(\xv', t),
\ee 
is path dependent and this relation cannot be used to construct $\Phi_\theta(\xv, t)$ unambiguously. 
Specialising to two dimensions, while introducing a third dimension for convenience (but no $z$ dependence), we can decompose the score in a gradient and a curl,
\be
\sv_\theta(\xv, t)  = \nabla \Phi_\theta(\xv, t)  +\nabla\times \Av_\theta(\xv, t)  \qqquad \Av_\theta(\xv, t)  = \left(0,0,A_\theta(\xv, t) \right),
\ee
or
\be
s_{x,\theta}(\xv, t)  = \partial_x \Phi_\theta(\xv, t)  + \partial_y A_\theta(\xv, t) , 
\qqquad
s_{y,\theta}(\xv, t)  = \partial_y \Phi_\theta(\xv, t)  - \partial_x A_\theta(\xv, t) .
\ee
Both scalar functions satisfy a Poisson equation 
\be
\nabla^2\Phi_\theta(\xv, t)  = \nabla\cdot \sv_\theta(\xv, t),
\qqquad
\nabla^2A_\theta(\xv, t)  = -\left(\nabla\times \sv_\theta(\xv, t) \right)_z,
\ee
sourced by derivatives of the score. 
Solutions depend crucially on the boundary conditions. 
However, at the boundaries of the data manifold, the score is not well determined, due to the lack of data to train the model well. These Poisson equations also make clear the possibility of adding a harmonic function $h(\xv)$, with $\nabla^2 h(\xv)=0$, to either, without affecting the score.
We emphasise that these observations do not undermine score-based diffusion models, which only require knowledge of the score itself, but complicate the explicit construction of the data distribution from the trained score.

\subsection{Energy-based models}

If we insist on learning the data distribution directly, it is useful to turn to energy-based models, see e.g.\ Refs.~\cite{shouldEBMs, horvat2024gaugefreedomconservativityintrinsic,du2024reducereuserecyclecompositional}, and impose
\be
E_\theta(\xv,t )\approx -\log p(\xv,t),
\ee
directly, with an undetermined normalisation.
To build on the experience of score-based models, Ref.~\cite{du2024reducereuserecyclecompositional} proposed to use 
the following energy function,
\be
\label{eq:EBM}
E_\theta(\xv,t) = \half \bignorm \vv_\theta(\xv,t) \bignorm^2 \approx -\log p(\xv,t),
\ee
leading to the approximate score 
\be
-\partial_i E_\theta(\xv,t) = -  \vv_\theta(\xv,t)\cdot  \partial_i\vv_\theta(\xv,t)  \approx \partial_i\log p(\xv,t),
\ee
which is conservative by construction. 
The advantage of this formulation is that it can be used in the loss function (\ref{eq:loss}), with training proceeding in the standard manner. It is noted that the choice of energy parametrisation is not unique and different choices may lead to different loss manifolds during training, requiring a separate treatment when choosing hyperparameters.
Since the energy parametrisation $E_\theta(\xv,t)$ is differentiated, it is important to ensure it is smooth enough by a convenient selection of activation function/nonlinearity such as SiLU or Mish, rather than ReLU or LeakyReLU. The need for  additional derivatives makes this method more expensive. Finally, we note that the energy is semi-positive; due to the undetermined normalisation, this is not a restriction. 

Below we refer to the two approaches as score-based models (SBMs) and energy-based models (EBMs). We are in particular interested in the learned scores at the end of the backward process, which we denote as
\be
\label{eq:SBM-EBM}
\begin{aligned}
\mbox{SBM:} \quad & \quad \sv_\theta(\xv) = \lim_{t\to 0} \vv_\theta(\xv,t), \\
\mbox{EBM:} \quad & \quad  \sv_\theta(\xv) = -\nabla E_\theta(\xv) =  -\lim_{t\to 0} \nabla E_\theta(\xv,t) =  -\lim_{t\to 0} \vv_\theta(\xv,t) \cdot\nabla\vv_\theta(\xv,t).
\end{aligned}
\ee
Here $\vv_\theta(\xv,t)$ is the trained neural network, with (near-)optimal parameters $\theta^*$. Below we drop the $*$ from $\theta^*$ to avoid a cluttering of symbols. 

In the numerical implementation of the SBMs and the EBMs we used a time-conditioned feed-forward neural network, 
incorporating the $\mathbf{Z}_2$ symmetry of the target distribution (see below) and employing an exponential moving average (EMA) of the weights. More details can be found in Appendix~\ref{sec:comp}.

\section{Complex-valued quartic model}
\label{sec:quartic}

We now apply the diffusion models of the previous section to data generated using complex Langevin dynamics. Since the emphasis in this paper is on conceptual issues, we consider a simple and well-studied example with one degree of freedom. After complexification, $x\to z=x+iy$, the model has two degrees of freedom and hence we are interested in the distribution $p(x,y)$.
In Appendix~\ref{sec:gaussian} we verify that our findings are consistent in an exactly solvable {\em real-valued} model, in which the distribution $p(x,y)$ is known.

We consider the quartic model with a complex mass parameter \cite{Aarts:2013uza}
\be
 S = \half\sigma_0 x^2+\frac{1}{4}\lambda x^4, \qqquad\sigma_0 = A+iB.
\ee
Exact results can be obtained by a direct evaluation of the partition function,
\be
 Z = \int dx\, e^{-S(x)} = \sqrt{\frac{4\xi}{\sigma_0}}e^{\xi} K_{-\frac{1}{4}}(\xi),
\ee
 where $\xi=\sigma_0^2/(8\lambda)$ and $K_q(\xi)$ is the modified Bessel function of the second kind. Subsequently, moments $\mu_n = \E[x^n]$ are obtained by differentiating with respect to $\sigma_0$. Odd moments vanish.

\begin{figure}[t]
    \centering
    \includegraphics[width=0.49\linewidth]{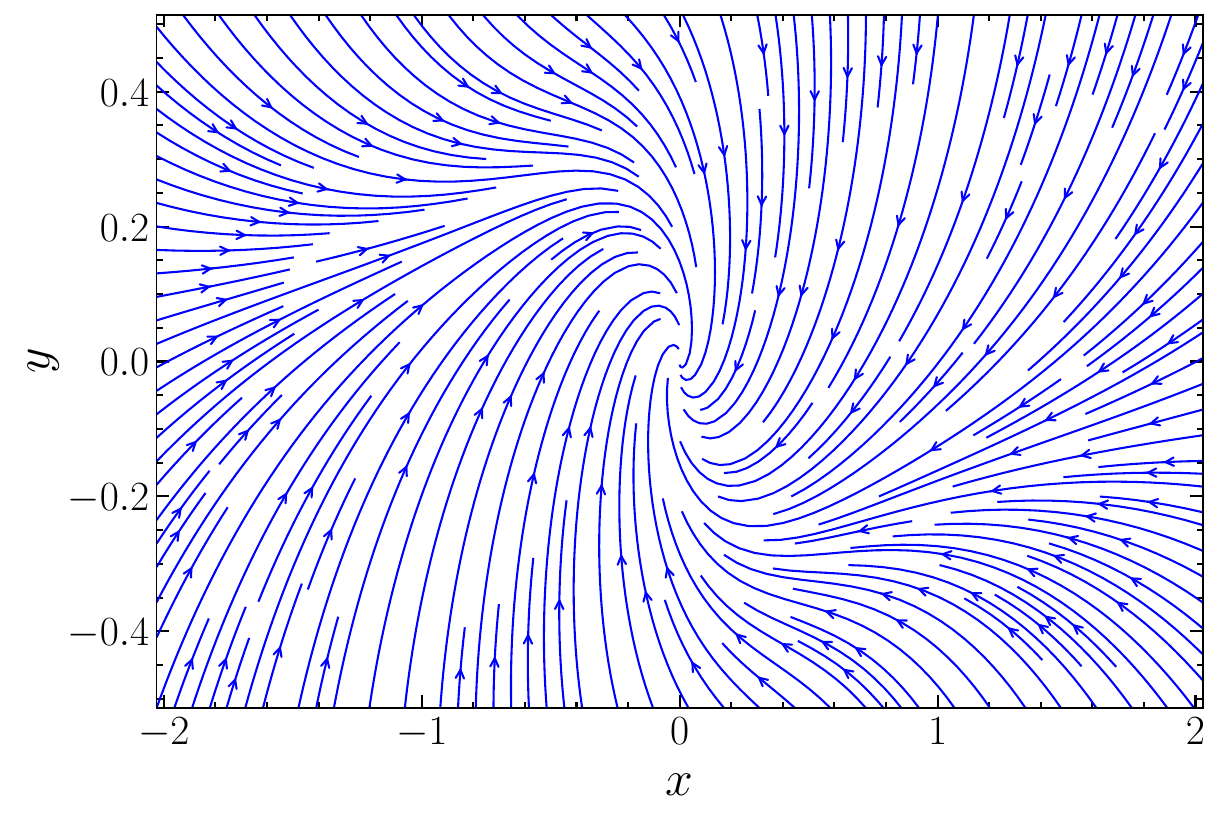}
    \includegraphics[width=0.49\linewidth]{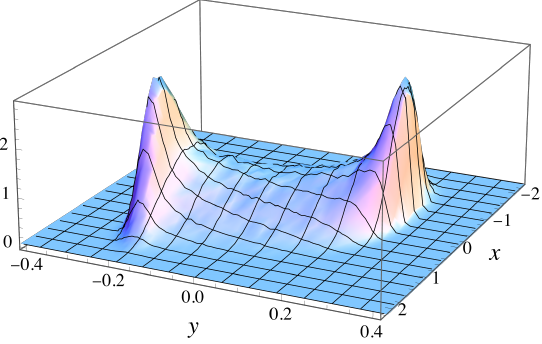}
    \caption{Complex-valued quartic model with parameters $\sigma_0=1+i$ and $\lambda=1$: CL drift in the complex plane (left) and histogram $P(x,y)$ obtained by sampling the CL process (right) \cite{Aarts:2013uza}.
     }
    \label{fig:CL-quartic}
\end{figure}

The model is of interest since it is possible to formulate exactly when CL dynamics works correctly and when it fails:
provided that $3A^2-B^2\geq 0$, the CL process is contained in a strip $-y_-<y<y_-$, with \cite{Aarts:2013uza}
\be
y_-^2 = \frac{A}{2\lambda}\left( 1- \sqrt{1-\frac{B^2}{3A^2}}\right), \qqquad y_->0.
\ee
In this case CL dynamics yields the correct results \cite{Aarts:2013uza}. 
When $3A^2-B^2<0$, the distribution is no longer contained and decays with a power law, $p(x,y)\sim (x^2+y^2)^{-3}$ for large $x,y$. Higher-order moments are then no longer well-defined and CL dynamics fails. In the numerical experiments we take $A=B=\lambda=1$, for which $y_-= 0.3029$. 
In Fig.~\ref{fig:CL-quartic} (left) we show the CL drift. The boundaries at $|y|=y_-$ are clearly visible: all arrows point inwards and hence the process cannot escape the strip $|y|<y_-$ in the case that real noise is used.

\begin{figure}[t]
    \centering
    \includegraphics[width=.49\linewidth]{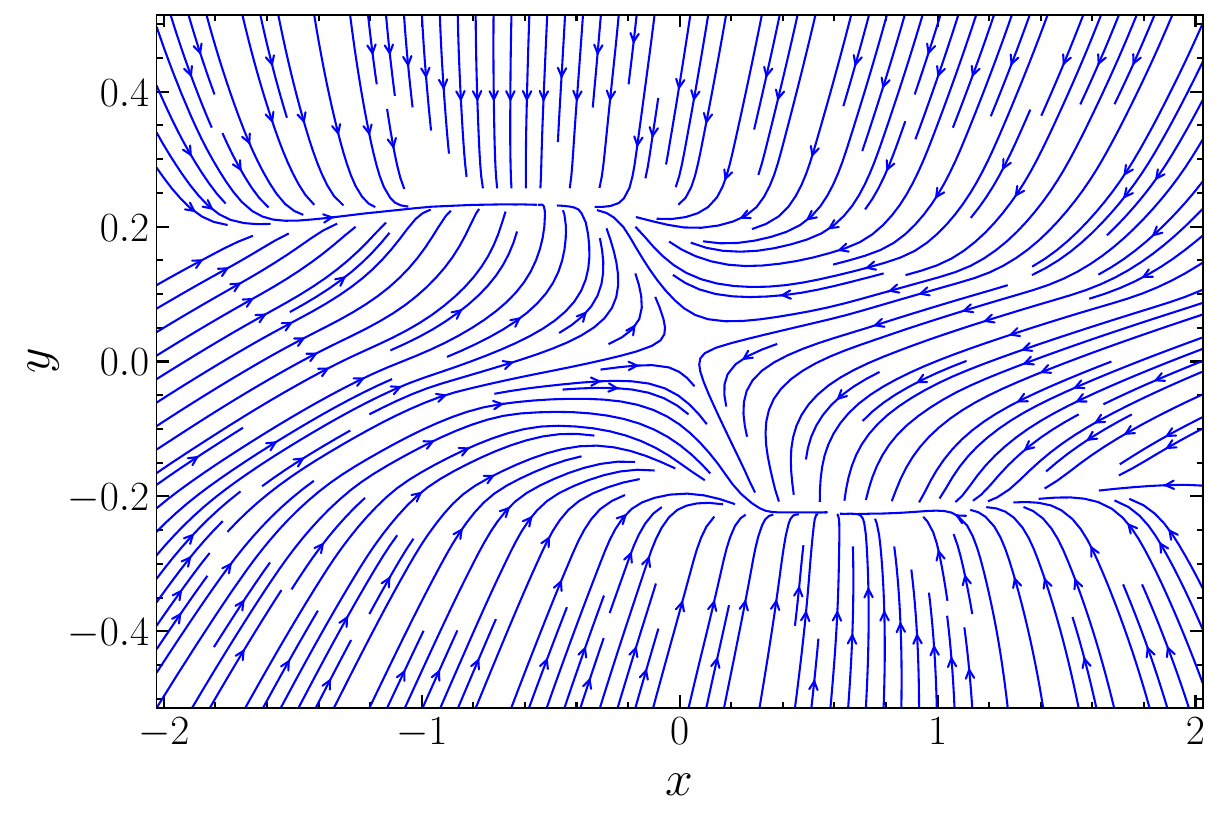}
    \includegraphics[width=.49\linewidth]{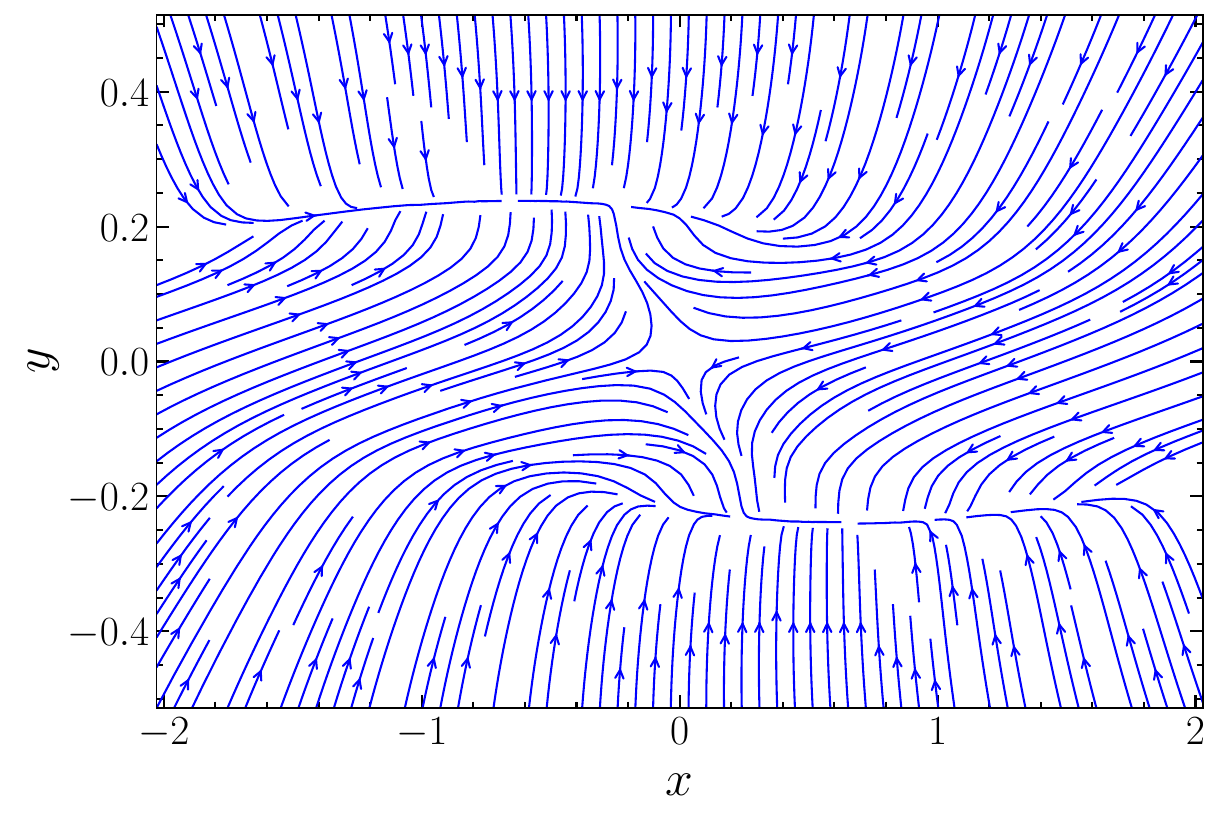}
    \caption{Learned scores in the quartic model at the end of the backward process, using a score-based (left) and energy-based (right) diffusion model. 
    }
    \label{fig:quartic-flows}
\end{figure}

We solve the CL process with real noise by discretising Eq.~(\ref{eq:CL}) with stepsize $\epsilon$ using the higher-order algorithm of Ref.~\cite{CCC:1987}, first applied to CL dynamics in Ref.~\cite{Aarts:2011zn}. This algorithm improves stepsize corrections from ${\cal O}(\eps)$ to ${\cal O}(\eps^{3/2})$. Collecting the data in a histogram yields the `empirical' histogram shown in Fig.~\ref{fig:CL-quartic} (right).  
As stated above, no analytical expression for this distribution is available.\footnote{In Ref.~\cite{Aarts:2013uza} an approximate solution of the FPE was given using a double expansion in terms of Hermite functions.} The distribution is strictly zero when $|y|> y_-$, consistent with the argument based on the drift, given above.

Without direct access to the distribution, it is hard to assess the reliability of the approach.
Diffusion models, however, can learn (the log-derivative of) this distribution. Notably, for a diffusion model it is irrelevant what the origin of the configurations is. 
We have trained the diffusion models using the CL data as the training set. Details can be found in Appendix~\ref{sec:comp}. 
In Fig.~\ref{fig:quartic-flows} we show the learned scores at the end of the backward process in the score-based (left) and energy-based (right) formulations, see Eq.~(\ref{eq:SBM-EBM}).
These vector fields should be contrasted with the CL drift of Fig.~\ref{fig:CL-quartic} (left). The first observation is that they are different, as they should be. Recall that the CL drift is used in the CL equation with noise in the $x$ direction only, whereas the score is used in the diffusion model with noise applied in both directions. Moreover, the CL drift is not integrable, since $\partial_x K_y(\xv) \neq \partial_y K_x(\xv)$. 
In particular it has an attractive fixed point at the origin, while the origin in the diffusion models is a saddle point. Reversely, in the diffusion models, the two peaks, at $\xv_{\rm peak}\sim \pm(0.6, -0.25)$, are attractive. 
To contain the data within the strip, also in the diffusion models the drift is pointing inwards at $|y| = y_-$. 
For completeness, we compare these drifts with Lefschetz thimbles in Appendix~\ref{sec:thimbles}.

\begin{figure}[t]
    \centering
    \includegraphics[width=.49\linewidth]{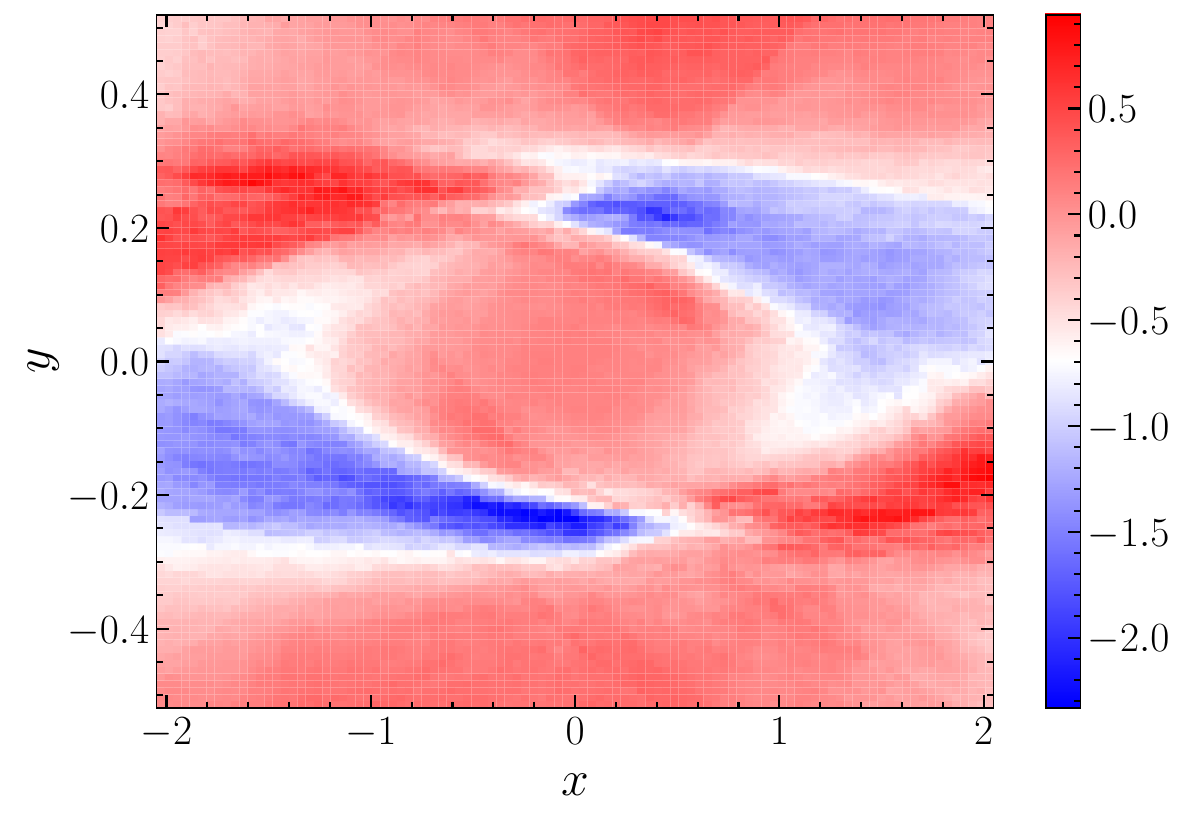}
    \includegraphics[width=.49\linewidth]{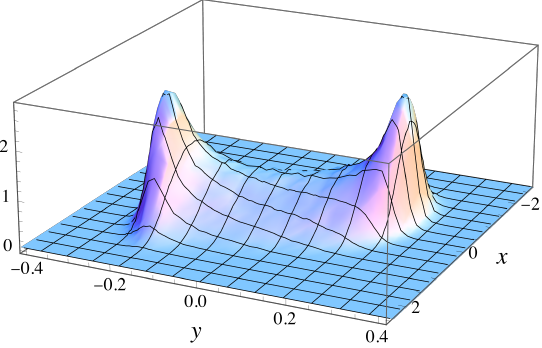}
    \caption{Quartic model using the score-based formulation: 
    Curl of the score, averaged over 10 independently trained models (left) and 
    histogram obtained by sampling data using the process learnt by the score-based model (right).
}
    \label{fig:quartic-score-based}
\end{figure}

\begin{figure}[t]
    \centering
    \includegraphics[width=.49\linewidth]{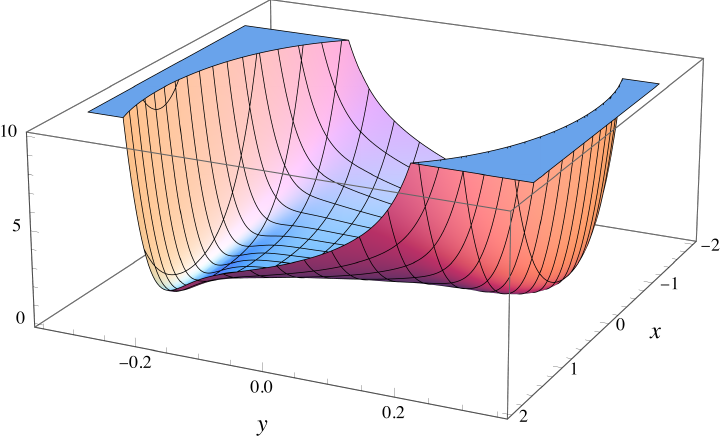}
    \includegraphics[width=.49\linewidth]{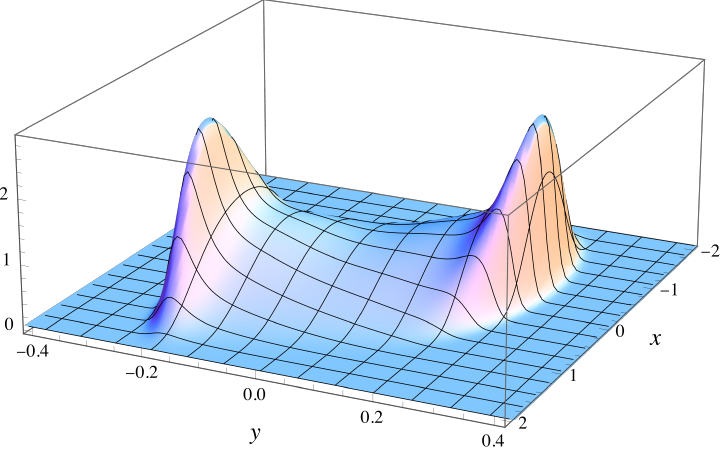}
    \caption{Quartic model: Energy $E_\theta(\xv)$ learned in the energy-based diffusion model 
    (left) and the corresponding distribution $p_\theta(\xv)\sim \exp[-E_\theta(\xv)]$ (right).
    }
    \label{fig:quartic-EBM}
\end{figure}

\begin{figure}[t]
    \centering
    \includegraphics[width=.45\linewidth]{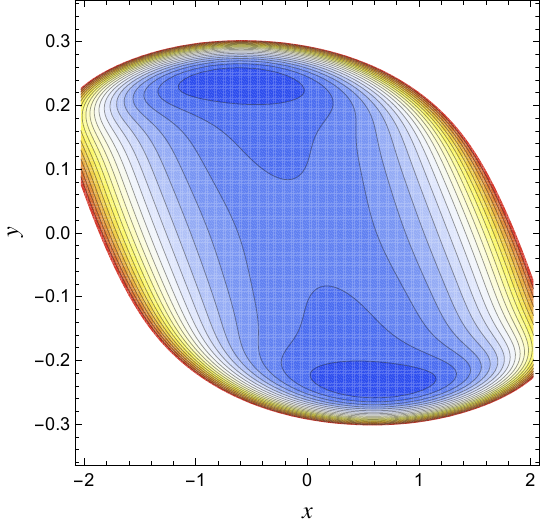}
    \caption{Quartic model: Contour plot of the energy learned in the energy-based model.
    }
    \label{fig:quartic-EBM-contour}
\end{figure}

While the two drifts in Fig.~\ref{fig:quartic-flows} look quite similar, the one obtained in the score-based formulation is not conservative. This is demonstrated in Fig.~\ref{fig:quartic-score-based} (left), where the curl of the score, $\partial_x s_y(\xv) - \partial_ys_x(\xv)$, is shown, averaged over 10 independently trained models.  Hence we cannot integrate the score directly. We can, however, obtain the distribution by sampling the process, as in the CL case. The result is shown in Fig.~\ref{fig:quartic-score-based} (right). Comparing the histograms in Figs.~\ref{fig:CL-quartic} and \ref{fig:quartic-score-based}, we observe that the trained score-based model captures the two peaks characteristic of this model as well as the boundary restrictions from the training data, although the edges at $|y|=y_-$ seem less sharp. We discuss a more quantitative comparison below. 

In the energy-based formulation, the energy $E_\theta(\xv)$ is learned directly. In Fig.~\ref{fig:quartic-EBM} (left) we show a three-dimensional plot, restricted to $E_\theta(\xv)\leq 10$ for clarity. In Fig.~\ref{fig:quartic-EBM-contour} we show the same energy using a contour plot. We note the two (shallow) minima around  $\xv_{\rm peak}$ and a quickly rising energy outside the main region of interest. The score is given by the gradient of the energy and is conservative by construction; we have verified that this is the case within numerical precision when the derivatives are computed numerically. 

In contrast to the cases above, we now have direct access to the distribution, by exponentiation and without any further sampling. The result is shown Fig.~\ref{fig:quartic-EBM} (right). We have determined the undetermined prefactor by normalising the distribution to 1. It should not come as a surprise that the distribution again resembles the ones obtained previously. The edges at $|y|=y_-$ seem as sharp as in the original CL distribution.
We emphasise that this is the first time a parametrisation of the distribution sampled by a CL process is obtained in a non-trivial case without explicitly constructing a histogram by further sampling.

\begin{figure}[t]
    \centering
    \includegraphics[width=.9\linewidth]{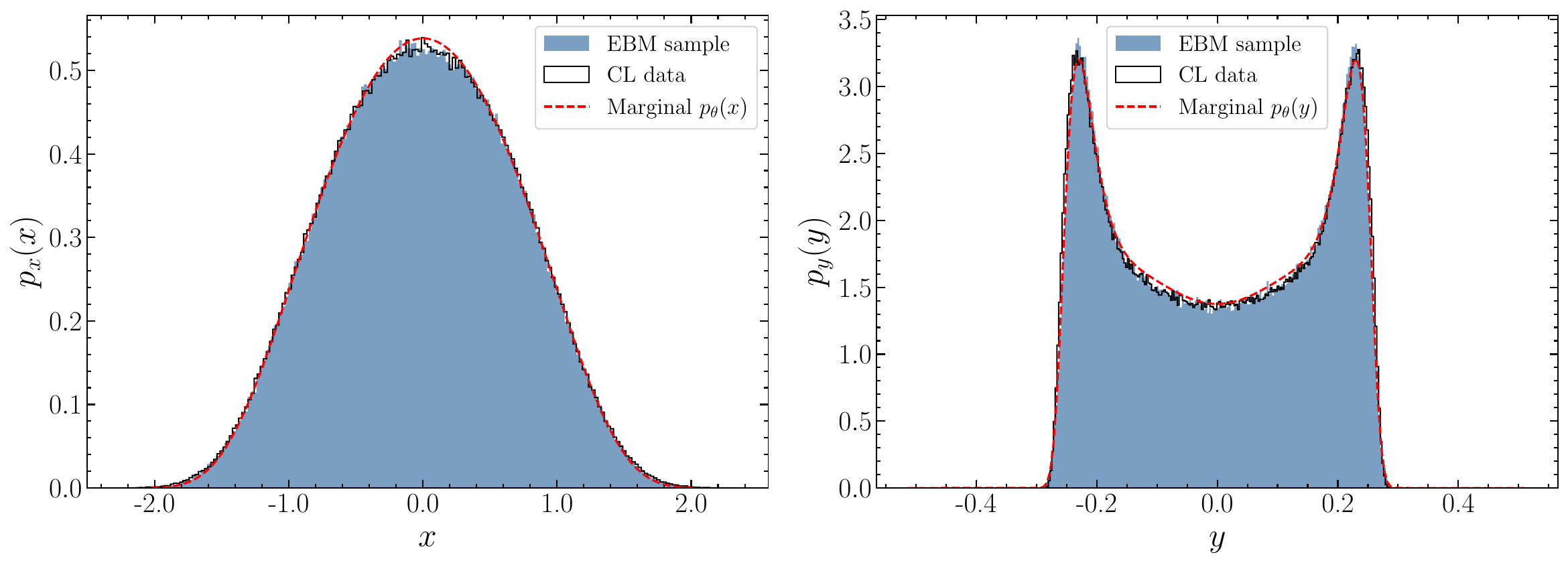}
    \caption{Marginalised distributions $p_x(x)$ and $p_y(y)$ in the quartic model obtained from $E_\theta(\xv)$ and by direct sampling using the CL process (data) and the energy-based model (EBM sample).
     }
    \label{fig:quartic-margin}
\end{figure}

To make a semi-quantitative comparison, we construct the marginal distributions
\be
p_x(x) = \int dy\, p_\theta(\xv), \quad p_y(y) = \int dx\, p_\theta(\xv), \quad p_\theta(\xv) = \frac{\exp[-E_\theta(\xv)]}{\int dxdy\, \exp[-E_\theta(\xv)]},
\ee
by numerical integration. In Fig.~\ref{fig:quartic-margin} we compare these distribution with those obtained by the CL process and by direct sampling of the energy-based model, collecting the data in histograms in the latter cases. We observe that the three distributions are in good agreement, with small deviations reflecting fluctuations in ensembles with a finite sample size.

\begin{table}[t]
    \centering
    \scalebox{0.9}{
    \begin{tabular}{c|ll|ll}
        \toprule
        & \; Re $\mu_2$ & \;\;\; Im $\mu_2$ & \; Re $\mu_4$ & \;\;\; Im $\mu_4$  \\
        \midrule
        Exact     &  0.428142      & $-$0.148010      &  0.423848        & $-$0.280132       \\
        CL        &  0.4281(5)     & $-$0.1481(2)     &  0.4232(11)      & $-$0.2798(6)      \\
        SBM       &  0.4259(2)(5)  & $-$0.1473(1)(3)  &  0.4237(4)(14)   & $-$0.2777(2)(9)     \\ 
        EBM       &  0.4264(1)(37) & $-$0.1487(1)(15) &  0.4192(2)(61)   & $-$0.2795(1)(39)  \\
        MCMC-EBM  &  0.4254(2)(75) & $-$0.1497(1)(31) &  0.4169(3)(122)  & $-$0.2802(2)(80)  \\
        \midrule
        & \; Re $\mu_6$ & \;\;\; Im $\mu_6$ & \; Re $\mu_8$ & \;\;\; Im $\mu_8$  \\
        \midrule
        Exact     &  0.580445         & $-$0.587746      &  0.95105      & $-$1.39336 \\
        CL        &  0.5787(26)       & $-$0.5866(18)    &  0.9482(87)   & $-$1.3901(84) \\
        SBM       &  0.594(1)(4)      & $-$0.5882(1)(3)  &  1.031(2)(14) & $-$1.435(3)(14) \\ 
        EBM       &  0.569(1)(12)     & $-$0.5834(4)(11) &  0.918(2)(26) & $-$1.374(2)(32) \\
        MCMC-EBM  &  0.565(1)(24)     & $-$0.584(1)(22)  &  0.912(2)(49) & $-$1.377(2)(61) \\
        \bottomrule
    \end{tabular}
    }
    \caption{Moments $\mu_n$ for the quartic model with parameters $\sigma_0=1+i, \lambda=1$, evaluated using complex Langevin (CL) dynamics, with statistical errors, score-based (SBM) and energy-based (EBM) diffusion models, with statistical and systematic errors, and an MCMC computation employing the energy learned using the EBM, with statistical errors and systematic errors.
    }
    \label{tab:quartic_moments}
\end{table}

\begin{table}[t]
    \centering
    \scalebox{0.9}{
    \begin{tabular}{c|ll|ll}
        \toprule
        & \;\;\; Re $\kappa_2$ & \;\;\; Im $\kappa_2$ & \;\;\; Re $\kappa_4$ & \; Im $\kappa_4$  \\
        \midrule
        Exact       & \;\; 0.428142      & $-$0.148010      &  $-$0.060347      & 0.100083       \\
        CL          & \;\; 0.4280(5)     & $-$0.1480(2)     &  $-$0.0606(6)     & 0.1003(5)      \\
        SBM         & \;\; 0.4259(2)(6)  & $-$0.1473(1)(3)  &  $-$0.0554(2)(4)  & 0.0986(1)(4)  \\
        EBM         & \;\; 0.4273(2)(9)  & $-$0.1478(1)(2)  &  $-$0.0607(2)(3)  & 0.1001(1)(4)  \\
        MCMC-EBM    & \;\; 0.4254(2)(76) & $-$0.1497(1)(31) & $-$0.0602(2)(44) & 0.1030(1)(62)  \\
        \midrule
        & \;\;\; Re $\kappa_6$ & \;\;\; Im $\kappa_6$ & \;\;\; Re $\kappa_8$ & \; Im $\kappa_8$  \\
        \midrule
        Exact       &  $-$0.00934       & $-$0.19222       & \;\; 0.41578      & 0.5923 \\
        CL  &  $-$0.009(1)      & $-$0.194(2)      & \;\; 0.414(5)     & 0.60(1) \\
        SBM &  $-$0.0131(4)(7)  & $-$0.1863(6)(11) & \;\; 0.423(2)(6)  & 0.557(3)(4) \\
        EBM         &  $-$0.0102(4)(6)  & $-$0.193(1)(2)   & \;\; 0.422(2)(5)  & 0.594(3)(8) \\
        MCMC-EBM    &  $-$0.0124(4)(32) & $-$0.205(1)(20) & \;\; 0.468(2)(51) & 0.661(4)(81) \\
        \bottomrule
    \end{tabular}
    }
    \caption{As in Table \ref{tab:quartic_moments}, for the cumulants $\kappa_n$.}
    \label{tab:quartic_cumulants}
\end{table}

To make a final quantitative comparison, we evaluate moments  $\mu_n$ and cumulants $\kappa_n$, starting from
\be
\mu_n = \E[z^n] = \int dz\, \rho(z) z^n  \stackrel{?}{=}   \int dxdy\, p(x,y) (x+iy)^n.
\ee
Cumulants follow in the usual way \cite{Aarts:2024rsl}.
 In Tables \ref{tab:quartic_moments} and \ref{tab:quartic_cumulants} we present the results for $\mu_n$ and $\kappa_n$ respectively, with $n=2,4,6,8$.
In each case the first line is the exact result. The second line is obtained using CL sampling and includes a statistical error. We remind the reader that the CL process provides the training data and is therefore the benchmark for the comparison. 
 The third and fourth lines are obtained using trained score-based and energy-based diffusion models. To estimate the systematic uncertainty, we follow the procedure outlined in Appendix~\ref{sec:system}, creating an ensemble of $N=10$ independently trained models, with the same set of (near-)optimal hyperparameters.   
 The first (second) number between the brackets represents the statistical (systematic) uncertainty. The systematic uncertainty dominates, reflecting the stochastic nature of the training procedure. 
 Errors in the moments and cumulants are determined independently, using an equal number of bootstrap samples, and cumulants have been constructed from central moments.  
 
The final line is obtained by taking the energy parametrisation $E_\theta(\xv)$ in the energy-based model and perform a Metropolis-Hastings Markov Chain Monte Carlo (MCMC) simulation using the learned energy in the accept/reject criterion. 
Again we used 10 trained EBMs, and combine the outcomes to yield an estimate of the systematic uncertainty due to the learned energies.
We emphasise that this approach was not feasible before, as the probability distribution was not expressed in terms of an energy function  $E_\theta(\xv)$, which can be evaluated at arbitrary $\xv$ within the data domain. 
We note that in this approach, after creating the data set and training the energy-based model, new configurations are created without any reference to the CL dynamics or the diffusion model, i.e., only the energy function is required. This opens the door to apply improved MCMC algorithms to theories with a sign problems eventually and should be explored further. 

Overall we observe good agreement between the various methods, with fluctuations representing mostly systematic uncertainties. It goes without saying that since the original weight $\rho(z)$ is complex-valued, we cannot employ a Metropolis step based on the original theory, as in the Metropolis-adjusted Langevin algorithm (MALA) \cite{roberts1998optimal}.

\section{Summary and outlook}
\label{sec:out}

We have used diffusion models to study the distribution sampled by a complex Langevin process. Since this distribution is not known {\em a priori}, this provides a new perspective on understanding CL dynamics for theories with a sign problem. We have noted that both score-based (SBM) and energy-based (EBM) diffusion models can learn the score, i.e., the log derivative of the distribution, and can subsequently be used to generate additional configurations. However, in score-based models the learned score is not conservative. While this does not affect the generative power, it means that the distribution itself cannot be obtained by integration. 
We found that the CL drift and the SBM and EBM scores all differ (the latter only due to the non-conservative component in the SBM), but yield consistent values for observables.  

Energy-based models give direct access to the (unnormalised) distribution via  $p_\theta(\xv)\sim \exp[-E_\theta(\xv)]$. One possible application is a detailed study of properties of the distribution in particular in regions where CL encounters problems. A second application, which we have explored here, is to use the learned energy in an MCMC simulation. In such a setup, CL is used to provide training data and the diffusion model is used to learn the energy. Afterwards, an MCMC simulation is used to generate additional configurations, without reference to the original CL process or diffusion model. This setup should be explored further. It goes without saying that such an approach is only useful when the original CL process converges correctly, i.e., the diffusion models as employed here will not solve the sign problem when CL fails. 
So far we have explored these aspects in a simple model. The obvious next step is to extend this to field theories, for which all ingredients are in place.

\el
\noindent
{\bf Acknowledgements} --  
GA thanks KZ and his group for the kind hospitality at CUHK-Shenzhen during the completion of this work. This visit was 
supported in part by the Royal Society International Exchanges 2024 Global Round 2 IES\textbackslash R2\textbackslash 242215.
GA is further supported by STFC Consolidated Grant ST/X000648/1.
DEH is supported by the UKRI AIMLAC CDT EP/S023992/1.
LW thanks the DEEP-IN working group at RIKEN-iTHEMS for its support in the preparation of this paper.
LW is also supported by the RIKEN TRIP initiative (RIKEN Quantum), JSPS KAKENHI Grant No. 25H01560, and JST-BOOST Grant No. JPMJBY24H9.
KZ is supported by the CUHK-Shenzhen university development fund under grant No.\ UDF01003041 and UDF03003041, and Shenzhen Peacock fund under No.\ 2023TC0179.
We acknowledge the support of the Supercomputing Wales and AccelerateAI projects, which are part-funded by the European Regional Development Fund (ERDF) via Welsh Government. 

\noindent
{\bf Research Data and Code Access} --
The code and data used for this manuscript will be available in version 2 on the arXiv.

\noindent
{\bf Open Access Statement} -- For the purpose of open access, the authors have applied a Creative Commons Attribution (CC BY) licence to any Author Accepted Manuscript version arising.

\appendix
\renewcommand{\theequation}{\Alph{section}.\arabic{equation}}

\section{Computational details}
\label{sec:comp}

Training data is generated by solving the discretised Langevin process with stepsize $\epsilon$ using the higher-order algorithm of Ref.~\cite{CCC:1987}, first applied to CL dynamics in Ref.~\cite{Aarts:2011zn}. This algorithm improves stepsize corrections from ${\cal O}(\eps)$ to ${\cal O}(\eps^{3/2})$. For both the complex-valued quartic model and the real-valued Gaussian mixture, we have generated an ensemble of $10^6$ configurations for training, which are preprocessed by scaling to zero mean and unit variance.

In the diffusion model, we employ a variance-expanding scheme, in which the drift term in Eq.~(\ref{eq:forward_process}) is set to zero, $K(\xv, t) = 0$. Note that due to the complexification, we have two degrees of freedom to consider. We choose the diffusion coefficient $g(t) = \sigma^{t/T}$ and pick $\sigma = 10$ and $T=1$.
We choose to run the backward process using 1000 steps for $10^6$ trajectories to obtain samples. Our choice of hyperparameters is summarised in Table \ref{tab:model_training_hyperparams}. More details can be found in Ref.~\cite{Aarts:2024rsl}.

\begin{table}[h]
\centering
\begin{tabular}{ll|ll}
\toprule
\textbf{Hyperparameter}    & \textbf{Value}  SBM / EBM       & \textbf{Hyperparameter}    & \textbf{Value}         \\ \midrule
Layers                     & [64, 64, 64] / [64, 64]               & Learning Rate              & 1e-4                   \\
Time Embedding dims        & 256 / 256                    & Batch Size                 & 512                    \\
Activation Function        & Leaky ReLU / SiLU                   & Optimizer                  & Adam                   \\
Weight Initialization      & LeCun Uniform \cite{LeCun1998}       & Max Epochs   & 300                    \\ \bottomrule \\
\end{tabular}

\caption{Model and hyperparameters used in training the score-based (SBM) and energy-based (EBM) diffusion models for the complex-valued quartic theory. Weights with the best loss are saved during the training process and we employ early stopping. For the Gaussian mixture, the same setup is used, with two layers of 32 nodes and a batch size of 1024.
} 
\label{tab:model_training_hyperparams}
\end{table}

We denote by $\vv_\theta(\xv, t)$: $\mathbb{R}^d\times [0, T] \to \mathbb{R}^d$ a time-conditioned neural network for which the output has the same dimensions as the input data for every $t\in [0, T]$. In score-based diffusion models, $\vv_\theta(\xv,t) = \sv_\theta(\xv, t)$ directly, while in energy-based models, $\vv_\theta$ is a trainable function which parametrises an energy functional \cite{du2024reducereuserecyclecompositional} to approximate the log-likelihood of the target distribution, see Eqs.~(\ref{eq:EBM}, \ref{eq:SBM-EBM}).
Using this formulation we may follow a similar prescription in training the two schemes, while being careful with the choice of hyperparameters. In particular, since in energy-based models $E_\theta(\xv, t)$ is differentiated to obtain the score, $\vv_\theta(\xv, t)$ should be sufficiently smooth. Hence we have used the SiLU  activation function in EBMs and the LeakyReLU for the SBMs.  
Concerning the architecture for $\vv_\theta$, we employed time-conditioned feed-forward neural networks using a Gaussian Fourier feature mapping \cite{tancik2020:GaussianFourier} and incorporating the $\mathbb{Z}_2$ symmetry of the data/target distribution, following Ref.~\cite{mattheakis2020physicalsymmetriesembeddedneural}. 
For the weights, we keep exponential moving averages (EMA) \cite{Song:2020_improve}, which are used during sampling.

Finally, in the energy-based model, we perform a Metropolis-Hastings MCMC using the energy parametrisation $E_\theta(\xv, 0)$ in the accept/reject criterion. 
The update was tuned to $70-80\%$ acceptance rate, yielding about $10^6$ independent configurations. 
The same approach is followed for direct MCMC simulations in the Gaussian mixture.

\section{Real-valued Gaussian mixture}
\label{sec:gaussian}

To verify that our findings are not due to the complexification, we consider a real target distribution in two dimensions, namely the following Gaussian mixture, 
\be
p_0(\xv) = \half\left[ \cN\left(\xv;\xv_0, \sigma_0^2\Id\right) + 
\cN\left(\xv; -\xv_0, \sigma_0^2\Id\right) \right].
\ee
Note that $p_0(\xv)=p_0(-\xv)$, as in the case considered above. In the numerical experiments, we take $\xv_0=(x_0,y_0)=(1,-1)$ and $\sigma_0^2=1/16$ throughout. 
We use the real-valued Langevin drift 
\be
\label{eq:GK}
\Kv(\xv) = \nabla \log p_0(\xv) = -\frac{\xv}{\sigma_0^2} +\frac{\xv_0}{\sigma_0^2} \tanh\left(\frac{x_0x+y_0y}{\sigma^2}\right)
\ee
to generate configurations using Langevin dynamics.
Subsequently, we train the score-based and energy-based diffusion models on this data distribution and generate new configurations. 

\begin{figure}[t]
    \centering
    \includegraphics[width=.32\linewidth]{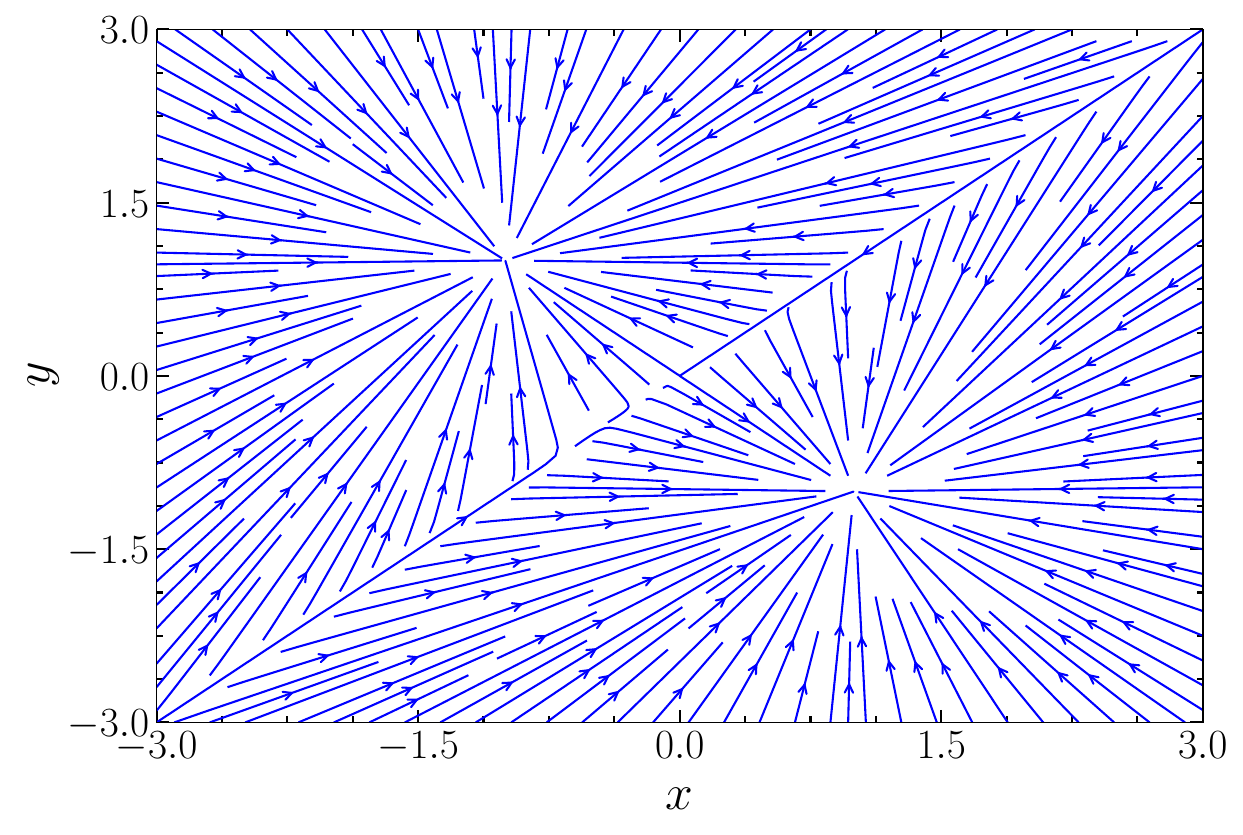}
    \includegraphics[width=.32\linewidth]{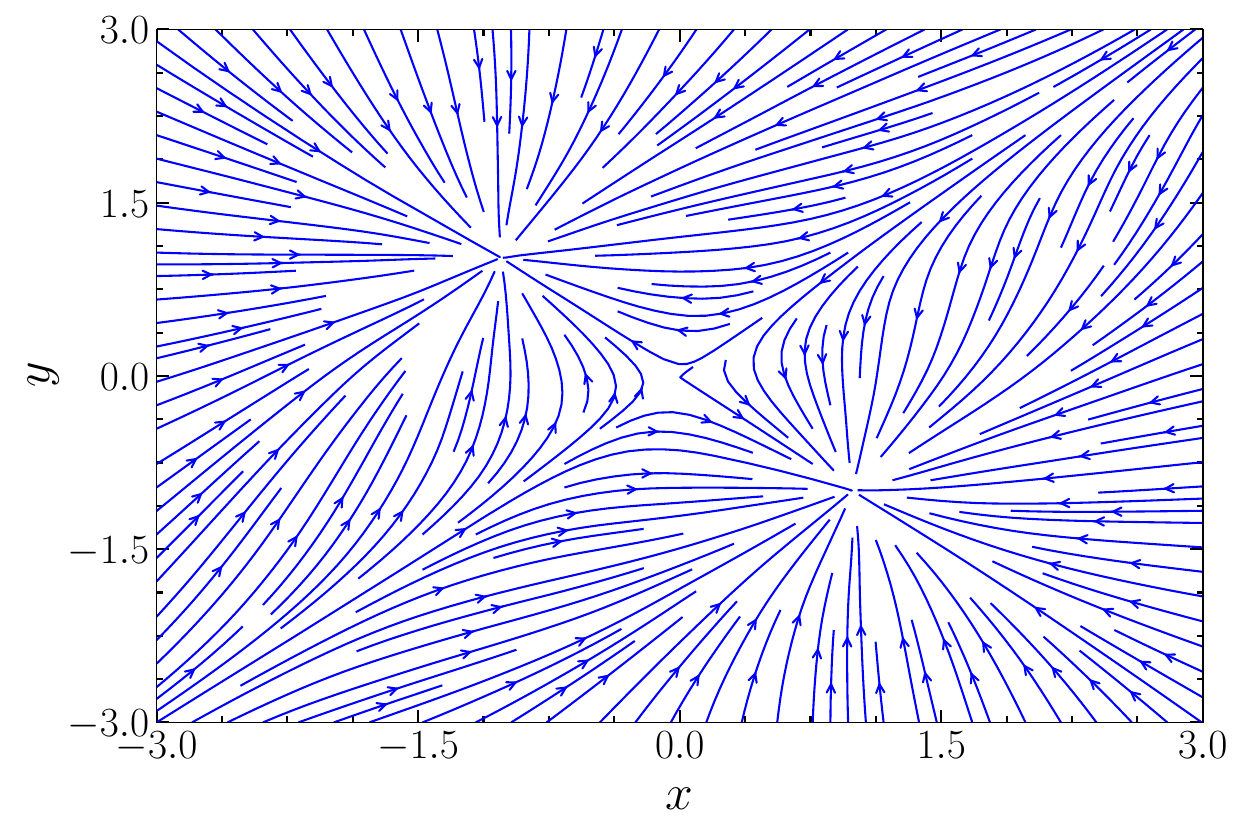}
    \includegraphics[width=.32\linewidth]{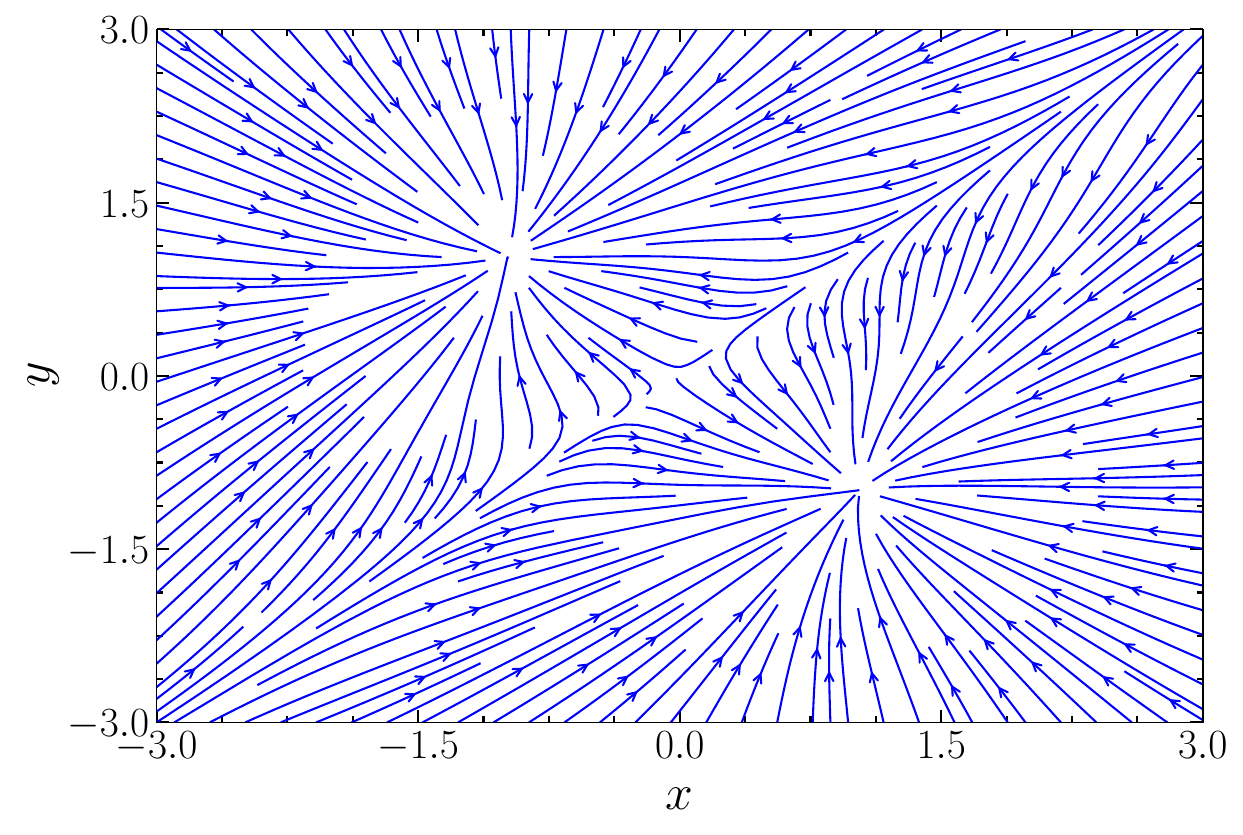}
    \caption{Score in the Gaussian mixture with parameters $\xv_0=(1,-1)$ and $\sigma_0^2=1/16$: exact (left), and using a score-based (middle) and energy-based (right) diffusion model.
    }
    \label{fig:Gflow}
\end{figure}
\begin{figure}[t]
    \centering
    \includegraphics[width=.32\linewidth]{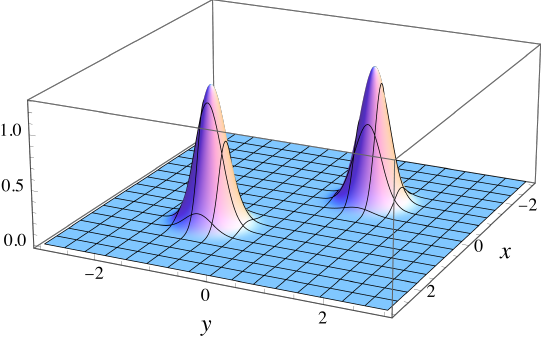}
    \includegraphics[width=.32\linewidth]{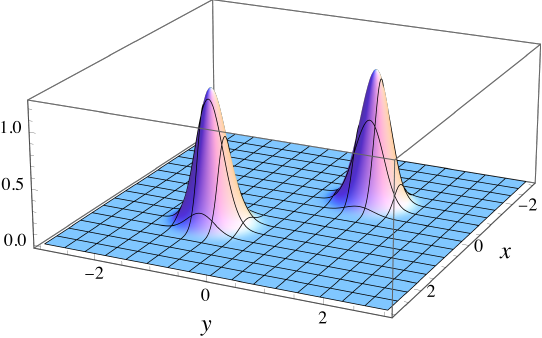}
     \includegraphics[width=.32\linewidth]{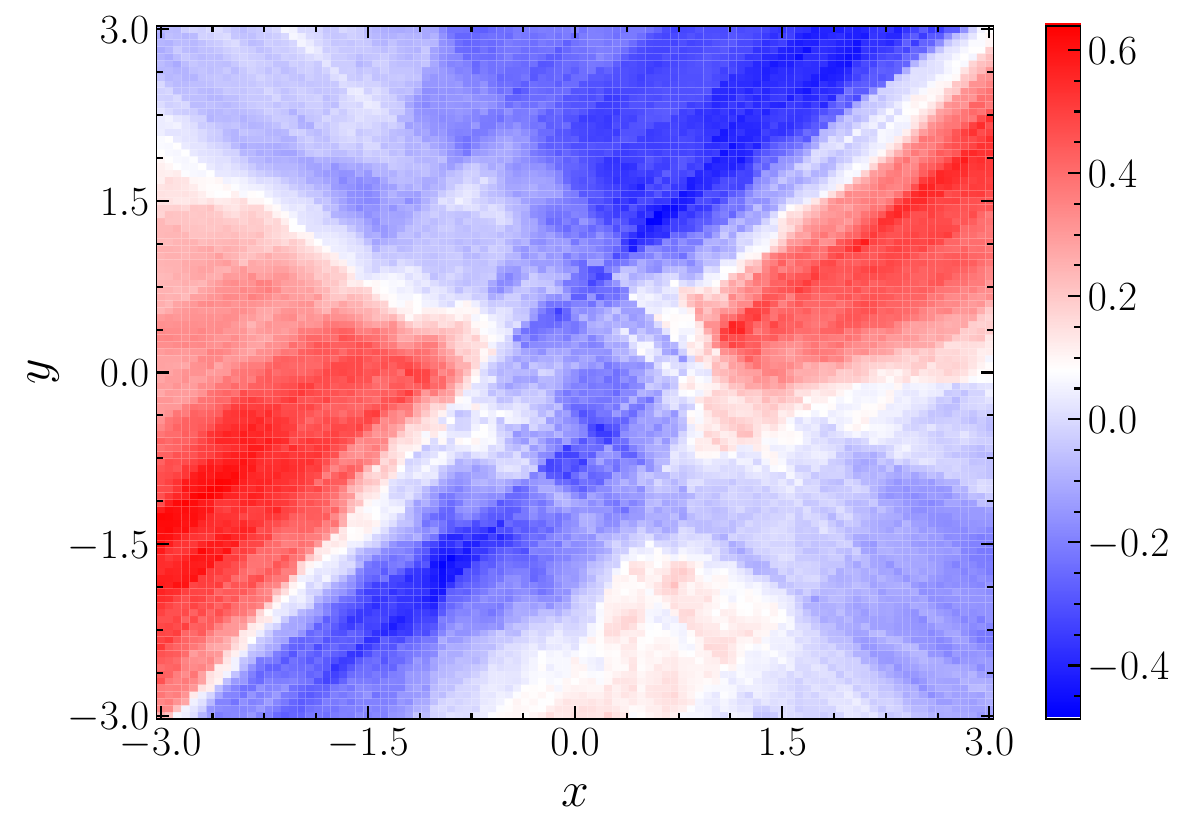}
    \caption{Gaussian mixture: exact distribution (left) and distribution obtained in the energy-based diffusion model (middle). 
    Curl of the score in the score-based formulation, averaged over 10 independently trained models. 
    }
    \label{fig:GPxy}
\end{figure}

In Fig.~\ref{fig:Gflow} we show the exact score (\ref{eq:GK}), and the approximate scores obtained by the score-based and energy-based diffusion models at the end of the backward process. Unlike in the case of the complex distribution, these vector fields have the same structure, as expected, with some deviations from the exact one visible by eye.
In Fig.~\ref{fig:GPxy} we present the exact distribution $p(\xv)$ (left) and the one obtained in the energy-based model (middle), by exponentiating the energy,  $p_\theta(\xv)\sim \exp[-E_\theta(\xv)]$, as before. The prefactor is obtained by normalising the distribution to 1 between the boundaries shown, i.e., $-3\leq x,y\leq 3$.
The score obtained in the score-based model is again non-conservative, and its curl is shown in Fig.~\ref{fig:GPxy} (right). Hence it cannot be integrated to obtain the distribution itself.

\begin{table}[t]
    \centering
    \scalebox{0.84}{
    \begin{tabular}{c|ll|ll}
       \toprule
                & \;\;$\bra x^2\ket$ & \;\;$\bra y^2\ket$ & \;\;\;\;$\bra x^4\ket$ & \;\;\;\;$\bra y^4\ket$   \\
        \midrule
        Exact       & 1.0625        & 1.0625        & 1.386719         & 1.386719          \\
        Data        & 1.0620(5)     & 1.06215(5)    & 1.3860(13)       & 1.3846(13)  \\
        SBM         & 1.0632(2)(7)  & 1.0638(2)(10) & 1.3954(4)(27)    & 1.3974(4)(28)  \\
        EBM         & 1.0621(2)(9)  & 1.0621(2)(6)  & 1.3846(4)(22)    & 1.3849(4)(15)   \\
        MCMC-EBM    & 1.0721(2)(122)& 1.0762(2)(155)& 1.4237(4)(310)   & 1.4350(4)(438)   \\
        MCMC-MH     & 1.0630(5)     & 1.0620(4)     & 1.3886(14)       & 1.3861(12)   \\
        \midrule
         & \;\;$\bra x^6\ket$ & \;\;$\bra y^6\ket$ & \;\;\;\;$\bra x^8\ket$ & \;\;\;\;$\bra y^8\ket$   \\
        \midrule
        Exact       &  2.11694      & 2.11694       & 3.67445        & 3.67445 \\
        Data        &  2.116(4)     & 2.113(3)      & 3.674(8)       & 3.667(9) \\
        SBM         &  2.149(1)(8)  & 2.155(1)(7)   & 3.778(3)(21)   & 3.792(3)(19) \\
        EBM         &  2.110(1)(5)  & 2.111(1)(4)   & 3.655(3)(13)   & 3.659(3)(10) \\
        MCMC-EBM    &  2.218(1)(71) & 2.246(1)(96)  & 3.933(3)(161)  & 4.003(3)(231)) \\
        MCMC-MH     &  2.122(3)     & 2.116(4)      & 3.692(8)       & 3.672(8) \\
        \bottomrule
    \end{tabular}
    }
%\end{table}
\vspace*{0.2cm}
%\begin{table}[t]
\vspace*{0.2cm}
    \centering
    \scalebox{0.84}{
    \begin{tabular}{c|l|l|l|l|l}
       \toprule
                & \;\;$\bra xy\ket$ & \;\;$\bra x^2y^2\ket$ & \;\;\;\;$\bra x^3y + xy^3\ket$ & \;\;\;\;$\bra x^3y^3\ket$ & $\bra x^4y^2 + x^2y^4\ket$  \\
        \midrule
        Exact       & $-$1             & 1.12891       & $-$2.375            & $-$1.41016        & \;\;2.94678  \\
        Data        & $-$1.0007(4)     & 1.1306(8)     & $-$2.3785(17)       & $-$1.4133(15)     & \;\;2.9535(35)    \\
        SBM         & $-$0.9997(1)(6)  & 1.1327(3)(16) & $-$2.3848(5)(34)    & $-$1.4238(5)(36)  & \;\;2.977(1)(7)   \\
        EBM         & $-$1.0000(1)(6)  & 1.1285(3)(12) & $-$2.3737(5)(24)    & $-$1.4087(5)(21)  & \;\;2.943(1)(5)   \\
        MCMC-EBM    & $-$1.0082(1)(116)& 1.1543(3)(255)& $-$2.4331(5)(558)   & $-$1.4652(5)(483) & \;\;3.066(1)(102)   \\
        MCMC-MH     & $-$0.9999(3)     & 1.1290(7)     & $-$2.3755(17)       & $-$1.4110(16)     & \;\;2.9492(31)    \\
        \midrule
         & \;\;$\bra x^5y + xy^5\ket$ & $\bra x^4y^4\ket$  &  \;\;$\bra x^5y^3 + x^3y^5\ket$ & \;\;\;\;$\bra x^6y^2 + x^2y^6\ket$ & $\bra xy^7 + x^7y\ket$ \\
        \midrule
        Exact       & $-$3.36719   	&  1.92299          & $-$3.99854        & \;\;4.4985           & $-$5.49658 \\
        Data        & $-$3.373(4) 	&  1.928(3)           & $-$4.009(7)       & \;\;4.510(8)         & $-$5.509(9)\\
        SBM         & $-$3.409(1)(9) 	&  1.960(1)(7)    & $-$4.076(2)(15)   & \;\;4.593(2)(19)     & $-$5.630(3)(25)\\
        EBM         & $-$3.361(1)(5)  	&  1.920(1)(4)    & $-$3.991(2)(8)    & \;\;4.488(2)(8)      & $-$5.479(3)(11)\\
        MCMC-EBM    & $-$3.521(1)(120) &  2.031(1)(87)  & $-$4.228(2)(181)  & \;\;4.776(2)(209)    & $-$5.876(3)(266)\\
        MCMC-MH     & $-$3.370(4)  	&  1.925(3)          & $-$4.004(6)       & \;\;4.504(7)         & $-$5.506(8)\\
        \bottomrule
    \end{tabular}
    }
     \caption{Various moments for the Gaussian mixture with parameters $\sigma_0=1/4, \mu_0=1$. As in Table \ref{tab:quartic_moments}, with the addition of the final line, using direct MCMC Metropolis-Hastings sampling from the target distribution, with statistical error.
    }
    \label{tab:gaussian_moments}
\end{table}

In Table \ref{tab:gaussian_moments} we show the results for the moments, observing good agreement. In this case it is possible to perform an MCMC simulation directly from the real probability distribution $p_0(\xv)$, allowing a comparison with the MCMC simulation based on the EBM. These results are given in the final line.

\section{Lefschetz thimbles}
\label{sec:thimbles}

\begin{figure}[t]
    \centering
    \includegraphics[width=.49\linewidth]{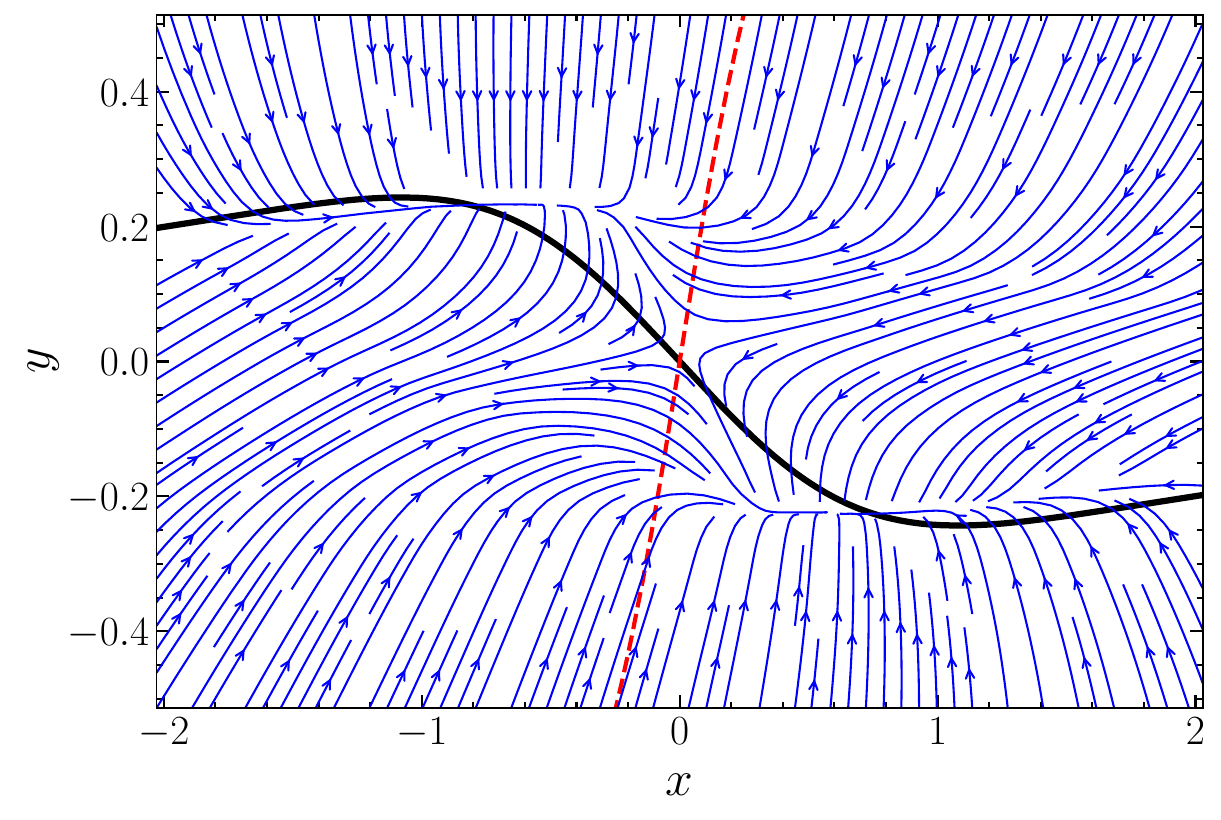}
    \includegraphics[width=.49\linewidth]{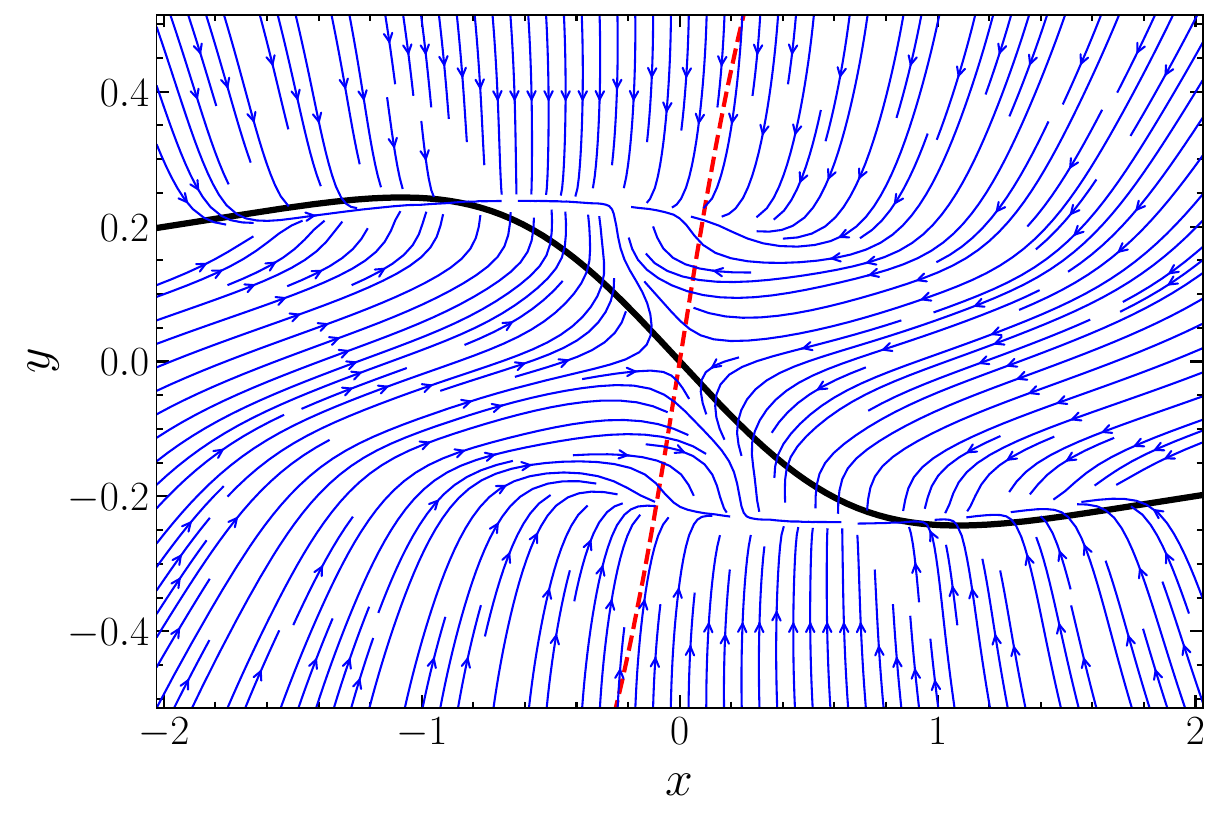}
    \caption{Stable (full) and unstable (thin dashed) thimbles superimposed on the drift determined in the score-based (left) and energy-based (right) diffusion models.
        }
    \label{fig:thimbles}
\end{figure}

Lefschetz thimbles are an alternative approach to enter the complexified manifold and potentially evade the sign problem \cite{Cristoforetti:2012su}. For the quartic model considered here, the stable and unstable thimbles were derived analytically in Ref.~\cite{Aarts:2013fpa}.  For completeness, we show in Fig.~\ref{fig:thimbles} the thimbles superimposed on the drifts obtained in the score-based and energy-based diffusion models.  We observe that the stable thimble follows in particular the attractive region near the top of the ridges. As already discussed in Ref.~\cite{Aarts:2013fpa}, in the thimble case the component of the drift in the imaginary direction is reversed and the critical point at the origin is attractive along the stable thimble. Hence the weight along the thimble peaks at the origin and not at the ridges.

\section{Estimating systematic effects}
\label{sec:system}

After determining the (near-)optimal choice of hyperparameters, we evaluate the uncertainty for the diffusion models using $N=10$ models, which are seeded differently and trained independently, but with the same choice of hyperparameters. Consider an observable $O$. For each model we estimate $O_i$, labelled by the index of the model, $i=1,..., N$, with some statistical uncertainty $\delta O_i$ determined using a jackknife analysis. The goal is to quantify the contribution of per-run statistical errors and estimate a (between-run) systematic error. 

For this we apply a random-effects meta-analysis known as DerSimonian-Laird \cite{DERSIMONIAN1986177,DERSIMONIAN2015139}. Each model is weighted by 
\be
  w_i = \frac{1}{\delta O_i^2},
  \ee
with a fixed-effect weighted mean, 
\be
  \overline O = \frac{\sum_i w_i O_i}{\sum_i w_i}. 
\ee
The excess scatter across models is quantified by a heterogeneity statistic such as Cochran's $Q$ \cite{Cochran1954101},
\be
  Q = \sum_i w_i \left(O_i - \overline O\right)^2.
\ee
A large $Q$ indicates variability beyond statistical deviation.

If we define the effective information quantity $C$ by
\be
C = \sum_i w_i - \frac{\sum_i w_i^2}{\sum_i w_i},
\ee
then the between-model variance is estimated by
\be
  \tau^2 = \max\left(0, \frac{Q - (N-1)}{C}\right).
\ee
Adjusting the weights for random effects,
\be
  w^*_i = \frac{1}{\delta O_i^2 + \tau^2}, 
\ee
 the adjusted weighted mean is 
\be
 O^* = \frac{\sum_i w_i^* O_i}{\sum_i w_i^*}.
\ee
Let 
\be
  a_i = \frac{w_i^*}{\sum_i w_i^*}.
\ee
The contributions to the error are then finally broken down as 
\begin{itemize}
    \item Statistical contribution: 
    $\sigma^2_{\mathrm{stat}} = \sum_i (a_i\,\delta O_i)^2$.
    \item Systematic contribution: 
    $\sigma^2_{\mathrm{sys}} = \tau^2\sum_i a_i^2$.
    \item Total random-effects uncertainty: 
    $\sigma^2_{\mathrm{total}} = 1/\sum_i w_i^*$.
\end{itemize}
In Tables \ref{tab:quartic_moments}, \ref{tab:quartic_cumulants} and \ref{tab:gaussian_moments} we quote $O^*(\sigma_{\mathrm{stat}})(\sigma_{\mathrm{sys}})$ for the score- and energy-based models.

\providecommand{\href}[2]{#2}\begingroup\raggedright\endgroup

%\nocite{*}
%\bibliographystyle{JHEP-arXiv}
%\bibliography{refs}

\end{document}